\documentclass[twocolumn,pre,superscriptaddress]{revtex4}   

\usepackage{dcolumn}
\usepackage{amsmath}
\usepackage{mdwlist}
\usepackage{calrsfs}

\usepackage{graphicx}
\usepackage{subfigure}
\usepackage{epstopdf}

\newcommand{\dd}{\mathrm{d}}

\newcommand{\e}{\mathrm{e}}
\newcommand{\half}{\tfrac12}
\newcommand{\set}[1]{\lbrace#1\rbrace}

\newcommand{\etal}{{\it{}et~al.}}
\newcommand{\defn}{\textit}

\newcommand{\mat}{\mathbf}

\begin{document}

\title{Random graph models for dynamic networks}

\author{Xiao Zhang}
\affiliation{Department of Physics, University of Michigan, Ann Arbor, MI
  48109}
\author{Cristopher Moore}
\affiliation{Santa Fe Institute, 1399 Hyde Park Road, Santa Fe, NM 87501}
\author{M. E. J. Newman}
\affiliation{Department of Physics, University of Michigan, Ann Arbor, MI 48109}
\affiliation{Santa Fe Institute, 1399 Hyde Park Road, Santa Fe, NM 87501}

\begin{abstract}  
  We propose generalizations of a number of standard network models, including the classic random graph, the configuration model, and the stochastic block model, to the case of time-varying networks.  We assume that the presence and absence of edges are governed by continuous-time Markov processes with rate parameters that can depend on properties of the nodes.  In addition to computing equilibrium properties of these models, we demonstrate their use in data analysis and statistical inference, giving efficient algorithms for fitting them to observed network data.  This allows us, for instance, to estimate the time constants of network evolution or infer community structure from temporal network data using cues embedded both in the probabilities over time that node pairs are connected by edges and in the characteristic dynamics of edge appearance and disappearance.  We illustrate our methods with a selection of applications, both to computer-generated test networks and real-world examples.
\end{abstract}

\maketitle

\section{Introduction}
\label{sec:intro}
Networked systems, such as social, technological, and biological networks, have been the subject of a vigorous research effort over the last decade~\cite{Newman10}, but most work has focused on static networks that do not change over time.  In reality, almost all networks do in fact change, with nodes or edges appearing or disappearing over time, and a body of new work aimed at quantifying, modeling, and understanding such temporal or dynamic networks has recently emerged, driven in part by the increasing availability of relevant data~\cite{HS12,Holme15}.

Data on dynamic networks comes in a variety of forms, but the most common form, and the one we consider in this paper, is that of a set of \defn{snapshots} of network structure taken at successive times, usually (though not always) evenly spaced.  Such sets are a special case of a more general ``multilayer'' or ``multiplex'' network, meaning a set of different networks defined on the same set of nodes~\cite{Boccaletti14,DGPA16}.  Multiplex networks include many non-dynamic kinds, such as social networks with different types of interactions between the same set of actors.  Our focus in this paper, however, is solely on dynamic networks.  We also limit ourselves to networks defined on a fixed and unchanging set of nodes, so that only edges appear and disappear, not nodes.  Our goal is to show how some of the most fundamental models for static networks can be generalized to the dynamic case and to demonstrate how comparisons between these models and real-world data can help us better understand the structure of the data.

Our models are built upon the assumption that the appearance and disappearance of network edges obeys a continuous-time Markov process.  That is, edges appear and disappear by making transitions from present to absent or \textit{vice versa} with fixed rates per unit time.  Crucially, however, these rates can differ from edge to edge and, as we will see, they can have quite complex structure.  If the rates of appearance and disappearance of edges are low compared to the rate at which we observe our snapshots of network structure, then consecutive snapshots will be correlated, a crucial feature of real dynamic networks.  In a friendship network, for instance, one expects to still be friends next week with most of the same people one is friends with this week, so there is a strong effect of previous friendship on future friendship probability, which is reproduced by our models.  In simple analyses of dynamic networks, researchers have in the past treated snapshots as independent measurements of network structure, analyzing each snapshot separately using conventional static network methods~\cite{Holme15}.  This, however, ignores the often strong correlations between snapshots and thereby also ignores a potential rich source of information hidden in the data.  For instance, it could be the case that two links in the network are each present in half the snapshots, but that one of these links flickers on and off rapidly, while the other one turns on and off more slowly; our Markov process model would distinguish these links as clearly different, while the more traditional model of analysis, ignoring correlations, would not.

An alternative way to think about our approach is that the fundamental unit of analysis in our calculations is not a single network but the entire \emph{history} of a network, and hence that the appropriate models are those that generate entire histories.  These are the models that we study in this paper.

Within this class of dynamic network models, we show how to formulate dynamic equivalents of the classic random graph, the configuration model, and the widely used stochastic block model, specifically its degree-corrected variant.  As we will show, dynamic variants of simple random graph models, for instance, allow us to properly define and measure the probability of an edge or the degree of a node in an evolving network, things that otherwise present a difficult moving target, and to attach values to the rate parameters that quantify how quickly edges appear and disappear.  Dynamic generalizations of block models allow us to infer large-scale structure, including (but not limited to) community structure, using maximum likelihood methods akin to those developed previously for the static case.

A number of other authors have previously considered dynamic generalizations of basic network models, particularly the stochastic block model~\cite{XFS10,Yang11,KL13,MM15,Ghasemian15,Xu14,MRV15,HXA15,SSTM15,XH13}.  The ordinary static version of the stochastic block model divides network nodes into groups or communities and then places edges between them with probabilities that depend on group membership.  Dynamic variants of this idea have been investigated in which nodes can change their community membership over time, which can cause edge probabilities also to change and hence edges to appear or disappear from one snapshot to the next.  Versions of this idea include the dynamic mixed-membership model of Xing~\etal~\cite{XFS10} and the multi-group membership model studied by Yang~\etal~\cite{Yang11} and Kim and Leskovec~\cite{KL13}.  In Matias and Miele~\cite{MM15} and Ghasemian~\etal~\cite{Ghasemian15}, group memberships can change but edges at successive times are independent conditioned on the groups.  Xu~\cite{Xu14} has studied a dynamic block model with edge dynamics controlled by a Markov process, which has some elements in common with our approach. Matias~\etal~\cite{MRV15} have considered ``longitudinal'' networks where contacts between nodes are governed by a Poisson process.

A little further from our focus in this paper are the multilayer stochastic block models studied for instance in Refs.~\cite{HXA15} and~\cite{SSTM15}.  As with dynamic models, these models generate a set of different networks or ``layers'' built upon the same set of nodes, but there is now no ordering of the layers or any assumption that adjacent layers are more similar than distant ones.  Han~\etal~\cite{HXA15} have used such multilayer models to derive more consistent estimation of community structure for certain data sets than those derived from standard stochastic block models.  Stanley~\etal~\cite{SSTM15} studied a variant in which different layers (``strata'' in their terminology) are generated from different underlying parameters.

In Section~\ref{sec:dynamic} we lay out the general principles behind our models, giving definitions and a variety of mathematical results for each of our models in turn.  In particular, we describe dynamic versions of three static models: the Erd\H{o}s--R\'enyi random graph, the configuration model of random graphs with a specific degree sequence, and the degree-corrected stochastic block model.  We also provide efficient algorithms for statistical inference using these models, showing how to perform a maximum-likelihood fit of each one to observed data.  In Section~\ref{sec:examples} we apply these models and algorithms to synthetic (i.e.,~computer-generated) test networks and to real-world examples, including technological and social networks.  In Section~\ref{sec:conclusions} we briefly describe our conclusions.

\section{Dynamic network models}
\label{sec:dynamic}
Each of the models we study has a fixed number~$n$ of nodes, plus edges between them that appear and disappear as the network evolves over time.  Starting from some initial condition at time~$t=0$, our models generate continuous-time network histories, where edges appear and disappear at a sequence of real-valued times.  In some data sets, events like these can be observed directly, for instance in a network of telephone calls where we are given the time and duration of each call.  Here, however, we assume that the network is only observed at a set of $T$ further snapshots, evenly spaced at integer times~$t=1,\ldots,T$.  Including the initial state there are, thus, a total of $T+1$ distinct snapshots.  Note, however, that the network is assumed to exist and to continue to evolve unobserved between the snapshots.

The fundamental idea behind all of the models we consider is that the edge between each node pair obeys a continuous-time Markov process, appearing and disappearing with constant rates, though the rates can differ from one node pair to another, depending on various latent properties of the nodes.  By choosing this dependence appropriately, we can model various kinds of dynamic network structure, including fluctuating density, degree distribution, or community structure.

To make our discussion more concrete, consider a particular pair of nodes in the network.  Let us define~$\lambda$ to be the rate (in continuous time) at which an edge appears between these two nodes where previously there was none, and let us define~$\mu$ to be the rate at which an existing edge disappears.  If we denote by $p_1(t)$ and $p_0(t)$ respectively the probabilities that there is and is not an edge between our nodes at time~$t$ then
\begin{align}
p_1(t+\dd t) &= p_1(t) + \lambda p_0(t) \>\dd t - \mu p_1(t) \>\dd t, \\
p_0(t+\dd t) &= p_0(t) - \lambda p_0(t) \>\dd t + \mu p_1(t) \>\dd t,
\end{align}
and hence $p_1$ satisfies the master equation
\begin{equation}
{\dd p_1\over\dd t} = - {\dd p_0\over\dd t} = \lambda p_0(t) - \mu p_1(t),
\end{equation}
which has the solution
\begin{equation}
p_1(t) = {\lambda\over\mu+\lambda} - c\,\e^{-(\mu+\lambda)t},
\label{eq:p1soln}
\end{equation}
where $c$ is an integration constant and we have made use of $p_0 = 1 - p_1$.

Now suppose that there is no edge between our two nodes at time~$t=0$, i.e.,~that $p_1(0)=0$, which corresponds to the choice $c = \lambda/(\mu+\lambda)$.  Then the probability of having an edge between our nodes at the next snapshot of the network, at time~$t=1$, is equal to $p_1(1)$, which takes the value
\begin{equation}
\alpha = {\lambda\over\mu+\lambda} \bigl[1 - \e^{-(\mu+\lambda)}\bigr].
\end{equation}
This is the probability of appearance of an edge between one snapshot and the next.  Similarly we can show that the probability of disappearance of an edge is
\begin{equation}
\beta = {\mu\over\mu+\lambda} \bigl[1 - \e^{-(\mu+\lambda)}\bigr].
\end{equation}
It will be more convenient to define our models in terms of per-snapshot probabilities such as these, which can always be calculated if necessary from the fundamental rates~$\lambda$ and~$\mu$.

\subsection{The dynamic random graph}
\label{sec:er_random}
The random graph~$G(n,p)$, famously studied by Erd\H{o}s and R\'enyi in the 1950s and 60s~\cite{ER59,ER60}, is perhaps the most fundamental of all network models.  In this model edges are placed between nodes pairs independently with probability~$p$ (or not with probability~$1-p$).  In this section we define the first and simplest of our dynamic network models as a direct dynamic counterpart to the random graph.

The definition of the model is straightforward.  Starting from some initial state at time~$t=0$, at every snapshot~$t$ each node pair not connected by an edge at the previous snapshot gains an (undirected) edge with probability~$\alpha$, or not with probability~$1-\alpha$.  Similarly each existing edge disappears with probability~$\beta$ or not with probability~$1-\beta$.  The net result after~$T$ time-steps is a sequence of $T+1$ snapshots which can be represented by a set of symmetric adjacency matrices~$\mat{A}(t)$ having elements $A_{ij}(t)=1$ if nodes~$i$ and~$j$ are connected by an edge in snapshot~$t$ and $A_{ij}(t)=0$ otherwise.

In the limit of long time $T\to\infty$, the average probability of an edge between two nodes in this model is given by Eq.~\eqref{eq:p1soln} to be $p=\lambda/(\lambda+\mu)=\alpha/(\alpha+\beta)$, the same for every node pair.  Hence the stationary distribution of the model is simply the random graph~$G(n,p)$.  It is in this sense that the model is a dynamic generalization of the random graph.

This is a particularly simple example of the class of models we study---we will look at more complex ones shortly---but even so there are various reasons to be interested in a model of this kind.  One could use it for instance to compute the time variation of network properties such as connectivity or component sizes, or the density of specific subgraphs---computations akin to the classic calculations of Erd\H{o}s and R\'enyi and others for the static case~\cite{ER59,ER60}.  Our primary interest in this paper, however, is in the use of this and other models as tools for understanding observed network data, using methods of statistical inference: we fit the model to the data by the method of maximum likelihood and the parameters of the fit tell us about our data in much the same way that fitting a straight line through a set of points can tell us about the slope of those points.

Suppose that we have a set of $T+1$ observed snapshots of some network, measured at uniform intervals over time.  If we hypothesize that the data were in fact generated from our dynamic random graph model, then the probability, or likelihood, that we observe this particular set of snapshots, given the parameters~$\alpha,\beta$ of the model, has the form
\begin{align}
P(\{ \mat{A}(t)\}| \,\alpha,\beta) &=
  \prod_{i<j} \Bigl[ P(A_{ij}(0)|\alpha,\beta) \nonumber \\
  &\qquad \times \prod_{t=1}^T
   P\bigl(A_{ij}(t)|\alpha,\beta,A_{ij}(t-1)\bigr) \Bigr].
\label{eq:likelihood_er}
\end{align}
Note that we have separate terms in this expression for the first snapshot and all succeeding snapshots.  The first snapshot differs from the others because it has no preceding snapshots and hence its probability is not conditioned on those before it.  The probabilities of all later snapshots, on the other hand, depend on the preexisting state of the network.  Because of the assumption that network evolution follows a Markov process, each snapshot only depends directly on the immediately preceding snapshot, hence the inclusion of~$A_{ij}(t-1)$ in the second product.

The two probabilities $P(A_{ij}(0)| \,\alpha,\beta)$ and $P\bigl(A_{ij}(t)| \,\alpha,\beta,A_{ij}(t-1)\bigr)$ are straightforward to write down.  The first, which represents the probability of observing~$A_{ij}(0)$ given no information about the previous history of the network, is equal to the stationary probability of an edge or non-edge within the model, which as we have said is $p=\alpha/(\alpha+\beta)$ for an edge, or $1-p$ for a non-edge.  Hence
\begin{align}
P(A_{ij}(0)| \,\alpha,\beta) &= p^{A_{ij}(0)} (1-p)^{1-A_{ij}(0)}.
\label{eq:stationary_er}
\end{align}
\begin{widetext}
The second probability is only a little more complicated, taking one of four values for edges that appear or not and ones that disappear or not:
\begin{equation}
 P(A_{ij}(t)| \,\alpha,\beta,A_{ij}(t-1)) = \alpha^{[1-A_{ij}(t-1)] A_{ij}(t)} 
(1-\alpha)^{[1-A_{ij}(t-1)][1-A_{ij}(t)]} 
 \beta^{A_{ij}(t-1) [1-A_{ij}(t)]} (1-\beta)^{A_{ij}(t-1) A_{ij}(t)}.
\label{eq:PAA_er}
\end{equation}
Substituting~\eqref{eq:stationary_er} and~\eqref{eq:PAA_er} into Eq.~\eqref{eq:likelihood_er} then gives us the full likelihood for our data.  In fact, as is often the case, it is more convenient to work with the logarithm~$\mathcal{L}$ of the likelihood, which has its maximum in the same place.  Taking the log of~\eqref{eq:PAA_er}, we have
\begin{align}
\mathcal{L} & = \log P(\{ \mat{A}(t)\}|\,\alpha,\beta)
  = \sum_{ij} \Bigl\lbrace A_{ij}(0)\log p + [1-A_{ij}(0)]\log (1-p) +
  \sum_{t=1}^T \Bigl[ [1-A_{ij}(t-1)] A_{ij}(t)\log \alpha \nonumber \\
  &\qquad + [1-A_{ij}(t-1)] [1-A_{ij}(t)] \log(1-\alpha) +A_{ij}(t-1) [1-A_{ij}(t)]\log\beta +A_{ij}(t-1)A_{ij}(t)\log(1-\beta) \Bigr]\Bigr\rbrace.
\label{eq:loglike_er}
\end{align}

Given the likelihood, we can estimate the parameters~$\alpha$ and~$\beta$ by maximizing, which gives
\begin{align}
\label{eq:alpha}
\alpha &= {\sum_{ij}\bigl[ A_{ij}(0) - p + \sum_{t=1}^T [1-A_{ij}(t-1)]A_{ij}(t) \bigr] 
\over \sum_{ij}\bigl[ A_{ij}(0) - p + \sum_{t=1}^T [1-A_{ij}(t-1)] \bigr]}, \\
\label{eq:beta}
\beta &= {\sum_{ij}\bigl[p - A_{ij}(0) + \sum_{t=1}^T A_{ij}(t-1)[1-A_{ij}(t)] \bigr] 
\over \sum_{ij}\bigl[p - A_{ij}(0) + \sum_{t=1}^T A_{ij}(t-1) \bigr]}.
\end{align}
\end{widetext}

Note that these expressions differ from the naive estimates of $\alpha$ and~$\beta$, given by the number of times an edge appeared or disappeared divided by the number of times it could potentially have done so.  The difference arises because the initial state of the network is chosen from the stationary distribution, and the probability $p$ that $A_{ij}(0)=1$ in this initial state itself depends on $\alpha$ and~$\beta$.  As $T \to \infty$ the effect of the initial state becomes progressively diluted relative to the effect of the other snapshots and Eqs.~\eqref{eq:alpha} and~\eqref{eq:beta} converge to the naive values.

Because $p$ appears on the right-hand side of \eqref{eq:alpha} and~\eqref{eq:beta}, calculating the rates~$\alpha$ and~$\beta$ requires us to find self-consistent solutions to the equations.  In fact, it is possible to eliminate the dependence on $p$ on the right-hand side and derive explicit closed-form equations, but the expressions are somewhat complicated.  In practice we have found it simpler just to solve Eqs.~\eqref{eq:alpha} and~\eqref{eq:beta} by iteration from a suitable initial condition.

What do these equations tell us?  For a given data set, they give us an optimal estimate---better than the naive estimate---of the rate at which edges appear and disappear in our network.  This gives us information about the correlation between adjacent snapshots.  The combined values of $\alpha$ and $\beta$ also give us the maximum-likelihood estimate of the average density of the network, via the average probability $p=\alpha/(\alpha+\beta)$ of an edge.

This model, however, while illustrative, is not, in practice, very useful.  Like the static random graph which inspired it, it is too simple to capture most of the interesting structure in real networks, and in particular it generates networks with Poisson degree distributions, wholly unlike those of real-world networks, which typically have broad and strongly non-Poisson distributions.  In the world of static network models, this latter shortcoming is remedied by the configuration model, a more sophisticated random graph that can accommodate arbitrary degree distributions~\cite{MR95,NSW01}.  In the next section, we show how to define a dynamic equivalent of the configuration model.

\subsection{Dynamic random graphs\\
with arbitrary expected degrees}
\label{sec:config}
The configuration model is a model of a random graph with a given degree sequence~\cite{MR95,NSW01}.  One fixes the degree~$d_i$ of each node~$i=1,\ldots,n$ and then places edges at random subject to the constraints imposed by the degrees.  This can be achieved in practice by endowing each node~$i$ with $d_i$ ``half-edges'' and choosing a matching of half-edges uniformly at random from the set of all possible matchings.  In the limit $n \to \infty$ the expected number of edges falling between nodes~$i$ and $j$ in this model is $d_id_j/2m$, where $m = \half \sum_i d_i$ is the total number of edges in the network, and the actual number of edges between each pair of nodes is Poisson distributed with this mean.  There is nothing in this model to stop a pair of nodes having two or more edges connecting them---a so-called \defn{multiedge}---and in general there will be some multiedges in networks generated using the configuration model.  Self-loops---edges connecting a node to itself---can and do also appear.  Although this is not realistic behavior for most real-world networks, versions of the configuration model that explicitly forbid multiedges and self-loops are much harder to work with than those that do not.  Moreover, if the degree distribution has finite mean and variance, the expected number of multiedges and self-loops in the network is constant, independent of~$n$, so they have vanishing density as $n \to \infty$.  For these reasons, one normally puts up with the presence of a few multiedges and self-loops for the sake of simplicity.

A commonly studied variant of the configuration model, which is easier to treat in some ways, involves explicitly placing between each node pair a  Poisson-distributed number of edges with mean $d_id_j/2m$.  In this variant, sometimes called the Chung--Lu model after two of the first authors to study it~\cite{CL02b}, the numbers of edges between node pairs are independent random variables, making analysis simpler.  The price one pays for this simplicity is that the degrees of individual nodes are no longer fixed, themselves being Poisson-distributed (and asymptotically independent) with mean~$d_i$.  Thus $d_i$ in this case represents not the actual degree but the expected degree of a node.  (The random graph of Erd\H{o}s and R\'enyi, with mean degree~$c$, is then the special case of this model where $d_i=c$ for all~$i$.)

In this section we define a dynamic analog of the Chung--Lu model in the sense of the current paper: its edges have a dynamics chosen so that the stationary distribution of the model is precisely the Chung--Lu model.  Since the Chung--Lu model can contain multiedges, we consider a process for adding and removing edges slightly different from the one of the previous section, such that each pair of nodes can have any nonnegative number~$k$ of edges connecting it.  Specifically, for each node pair we consider the Poisson process where edges are added at rate~$\lambda$, and each of the existing edges is removed independently at rate~$\mu$.  Thus $k$ is incremented with rate~$\lambda$, and decremented with rate~$k\mu$.  

Let $p_k(t)$ denote the probability that a node pair has $k$ edges at time~$t$.  Then $p_k$ satisfies the master equation
\begin{equation}
{\dd p_k\over\dd t} = \lambda p_{k-1}(t) + (k+1) \mu p_{k+1}(t)
                      - (\lambda + k \mu) p_k(t).
\label{eq:diffeq}
\end{equation}
We can solve this equation by defining a generating function $g(z,t)=\sum_{k=0}^\infty p_k(t)\,z^k$, multiplying both sides of~\eqref{eq:diffeq} by~$z^k$, and summing over~$k$ to get
\begin{equation}
{\partial g\over\partial t} = (z-1) \biggl[ \lambda g 
  - \mu {\partial g\over\partial z} \biggr].
\end{equation}
The general solution to this equation is
\begin{equation}
g(z,t) = \e^{\lambda(z-1)/\mu} f\bigl( (z-1) \e^{-\mu t} \bigr),
\label{eq:gzt}
\end{equation}
where $f(x)$ is any once-differentiable function of its argument satisfying $f(0)=1$, the latter condition being necessary to fulfill the normalization requirement $g(1,t)=\sum_k p_k(t) = 1$ for all~$t$.

In the limit of long time we have $g(z,t) \to \e^{\lambda(z-1)/\mu}$, which is the generating function of a Poisson-distributed variable with mean $\lambda/\mu$.  Hence the number of edges between any pair of nodes in this model is Poisson-distributed in the stationary state.  If we make the choice
\begin{equation}
\lambda = \mu {d_i d_j\over2m}
\label{eq:lambda}
\end{equation}
for some set of values~$d_i$, with $m = \half \sum_i d_i$ as previously and any value of~$\mu$, then the mean number of edges between nodes $i$ and~$j$ is $\lambda/\mu = d_i d_j/2m$.  In other words, the stationary state of this model is precisely the Chung--Lu model with expected degrees~$d_i$.

This then defines our model: to generate a dynamic network with $n$ nodes, we specify the expected degree~$d_i$ for each node and the parameter $\mu$.  We generate the initial state of the network from the Chung--Lu model with these expected degrees, and then generate future states by adding edges between each node pair $i,j$ at rate $\lambda_{ij} = \mu d_i d_j/2m$ and removing existing edges at rate~$\mu$.  We sample $T$ snapshots of the resulting network at integer intervals $t=1,\ldots,T$ which, along with the initial state at $t=0$, comprise the $T+1$ total snapshots generated by the model.  We represent these snapshots by adjacency matrices~$\mat{A}(t)$.

One could use this model for various purposes, such as making calculations of expected structural properties, but our principal interest here is again in fitting the model to observed network data.  As before we achieve this by maximizing a likelihood function, which has the same basic form as previously:
\begin{align}
P(\{ \mat{A}(t)\}|\{d_i\},\mu) &=
  \prod_{i<j} \biggl[ P(A_{ij}(0)|d_i,d_j,\mu) \nonumber \\
& \times \prod_{t=1}^T P\bigl(A_{ij}(t)|d_i,d_j,\mu,A_{ij}(t-1)\bigr) \biggr].
\label{eq:likelihood_cm}
\end{align}
The first probability on the right-hand side is straightforward to write down, since we know that the stationary distribution places a Poisson-distributed number of edges between nodes~$i$ and~$j$ with mean $d_i d_j/2m$.  Thus
\begin{equation}
 P(A_{ij}{(0)} | \{ d_i \},\mu)
  = { ( d_i d_j / 2m )^{A_{ij}(0)}\over A_{ij}(0)!} \e^{-d_i d_j /2m},
\label{eq:config}
\end{equation}
which is independent of~$\mu$.

The second probability $P(A_{ij}(t)|d_i,d_j,\mu,A_{ij}(t-1))$ is more involved, but the calculation is simplified by noting that even though the model can possess multiedges, the observed network data will normally have at most a single edge between any pair of nodes, so that the only allowed edge transitions are the appearance and disappearance of single edges.  

Suppose that a given node pair is connected by zero edges at time~$t=0$.  Then, setting $t=0$ in Eq.~\eqref{eq:gzt}, we find that $f(x) = \e^{-\lambda x/\mu}$, which implies that one timestep later at $t=1$ we have
\begin{equation}
g(z,1) = \e^{\lambda(z-1)(1-\e^{-\mu})/\mu} = \e^{(z-1)\beta d_i d_j/2m},
\end{equation}
where we have made use of Eq.~\eqref{eq:lambda} and for convenience defined the quantity
\begin{equation}
\beta = 1 - \e^{-\mu},
\end{equation}
which (by analogy with our use of the same symbol $\beta$ in Section~\ref{sec:er_random}) is equal to the total probability that an existing edge disappears during a single unit of time, i.e., between two successive snapshots.

The probabilities $p_{0\to0}$ and $p_{0\to1}$ of a transition from zero edges to, respectively, zero or one edges in a single timestep are then equal to the probabilities $p_0(1)$ and $p_1(1)$ of having zero or one edges at $t=1$.  These are given by the zeroth and first coefficients in the expansion of $g(z,1)$ in powers of~$z$:
\begin{align}
p_{0\to0} &= \e^{-\beta d_i d_j/2m}, \\
p_{0\to1} &= \beta\, {d_i d_j\over2m}\, \e^{-\beta d_i d_j/2m}.
\end{align}
By a similar method we also have
\begin{align}
p_{1\to0} &= \beta \e^{-\beta d_i d_j/2m}, \\
\label{eq:transition_config}
p_{1\to1} &= (1 - \beta) \e^{-\beta d_i d_j/2m}.
\end{align}
where we have ignored terms of second and higher order in $1/m$ in~\eqref{eq:transition_config}.

\begin{widetext}
  We can now write down the transition probability $P\bigl(A_{ij}(t)|d_i,d_j,\mu,A_{ij}(t-1)\bigr)$ as a function of $\beta$:
\begin{equation}
P\bigl(A_{ij}(t)|d_i,d_j,\beta,A_{ij}(t-1)\bigr)
  = (\beta d_i d_j/2m)^{[1-A_{ij}(t-1)] A_{ij}(t)} 
    \beta^{A_{ij}(t-1) [1-A_{ij}(t)]} 
    (1-\beta)^{A_{ij}(t-1) A_{ij}(t)} \e^{-\beta d_i d_j/2m}.
\end{equation}
Substituting this into Eq.~\eqref{eq:likelihood_cm} and taking logs, we get the following expression for the log-likelihood in our model:
\begin{align}
\mathcal{L} &= \sum_{ij} \biggl( A_{ij}(0)
  + \sum_{t=1}^T \bigl[ 1 - A_{ij}(t-1) \bigr] A_{ij}(t) \biggr) \log {d_i d_j\over 2m}
  + 2\bigl(m^{0\to1} + m^{1\to0}\bigr) \log \beta + 2m^{1\to1} \log{(1-\beta)}
  - 2m (1+T\beta),
\label{eq:loglike_config}
\end{align}
where
\begin{equation*}
m^{0\to1} = \frac{1}{2} \sum_{t=1}^T \sum_{ij} [1-A_{ij}(t-1)] A_{ij}(t)
\end{equation*}
is the total number of newly appearing edges in the observed data, and similarly
\begin{equation}
m^{1\to0} = \frac{1}{2} \sum_{t=1}^T \sum_{ij} A_{ij}(t-1) [1-A_{ij}(t)],\qquad
m^{1\to1} = \frac{1}{2} \sum_{t=1}^T \sum_{ij} A_{ij}(t-1) A_{ij}(t).
\end{equation}
Then, differentiating~\eqref{eq:loglike_config} with respect to~$\beta$, we find that the optimal value of~$\beta$ is the positive solution of the quadratic equation
\begin{equation}
mT \beta^2 - (mT + m^{0\to1} + m^{1\to0} + m^{1\to1} ) \beta
  + m^{0\to1} + m^{1\to0} = 0.
\label{eq:rho_config_max}
\end{equation}
Similarly, differentiating with respect to $d_i$ and bearing in mind that $m = \half \sum_i d_i$, we find that $d_i$ obeys
\begin{equation}
{2\over d_i} \sum_j \biggl[ A_{ij}(0) + \sum_{t=1}^T [1-A_{ij}(t-1)] A_{ij}(t) \biggr] - {1\over\sum_j d_j} \sum_{ij} \biggl[ A_{ij}(0) + \sum_{t=1}^T [1-A_{ij}(t-1)] A_{ij}(t) \biggr] - (1+T\beta) = 0,
\end{equation}
which has the solution
\begin{equation}
d_i = {1\over 1 + T\beta} \sum_j \biggl[ A_{ij}(0) + \sum_{t=1}^T [1-A_{ij}(t-1)] A_{ij}(t) \biggr].
 \label{eq:di_config}
\end{equation}
\end{widetext}

The sum in this expression is the number of edges initially connected to node~$i$ plus the number that later appear.  The divisor $1+T\beta$ is the effective number of independent measurements of an edge that we make during our $T$ snapshots.  If $\beta=0$, so that edges never appear or disappear, then in effect we only have one measurement of each edge---the initial snapshot at $t=0$.  Conversely, if $\beta=1$, so that every observed edge immediately disappears on the next snapshot, then all snapshots are independent and the number of independent measurements is $T+1$.  Thus Eq.~\eqref{eq:di_config} measures the number of observed edges between node pairs divided by the number of independent observations of each node pair.

Equations~\eqref{eq:rho_config_max} and~\eqref{eq:di_config} give us the maximum-likelihood estimates the rate parameter~$\beta$ and the expected degrees of the nodes.  We note two points:
\begin{enumerate}
\item These equations have to be solved self-consistently, since the first equation depends on $d_i$ via $m = \half \sum_i d_i$ and the second depends on~$\beta$.
\item Neither $\beta$ nor $d_i$ are equal to their naive estimates from the data.  One might imagine, for instance, that $d_i$ would be given by the average of $\sum_j A_{ij}(t)$ over all snapshots, but our results indicate that the maximum-likelihood estimate differs from this value.
\end{enumerate}
Both of these effects arise, as in the previous section, because of the information provided by the initial state.  Because the initial state is drawn from the stationary distribution, which depends on the model parameters, we can make a better estimate of those parameters by taking it into account than not.  On the other hand, the advantage of doing so dwindles as $T$ becomes large and vanishes in the $T\to\infty$ limit.

We could use these equations, for example, to define in a principled fashion an equivalent of the ``degree'' for a node in a dynamic network.  The actual degree of a node in such a network is a fluctuating quantity, but using our results one can define a single number~$d_i$ for each node that, like the degree in a static network, is a measure of the propensity of that node to connect to others.  We give some examples in Section~\ref{sec:examples}.

\subsection{Dynamic block models}
\label{sec:dsbm}
The stochastic block model is a random graph model of a network that incorporates modules or community structure---groups of nodes with varying densities of within- and between-group edges.  The standard stochastic block model, first proposed by Holland~\etal\ in~1983~\cite{HLL83}, is the community-structured equivalent of the random graph of Erd\H{o}s and R\'enyi, but like the latter it has shortcomings as a model of real-world networks because the networks it generates have Poisson degree distributions.  The degree-corrected stochastic block model~\cite{KN11a} is a variant on the same idea that is analogous to the model of Chung and Lu~\cite{CL02b}, allowing us to choose any set of values for the expected degrees of nodes, while also generating a community-structured network.  In this section we define a dynamic equivalent of the degree-corrected block model along similar lines to the models of previous sections and show how it can be used to infer community structure from dynamic network data.

The standard (static) degree-corrected block model divides a network of $n$ nodes into $k$ nonoverlapping groups labeled by integers $1,\ldots,k$.  Let us denote by $g_i$ the group to which node~$i$ belongs.  Then we place a Poisson-distributed number of edges between each node pair~$i,j$ with mean equal to $\omega_{g_ig_j} \theta_i \theta_j$, where $\theta_i$ is a degree-like parameter and $\omega_{rs}$ is a further set of parameters which control the density of edges within and between each pair of groups.  If the diagonal elements~$\omega_{rr}$ are greater than the off-diagonal ones, this model generates networks with conventional ``assortative'' community structure---dense in-group connections and sparser between-group ones---although other choices of~$\omega_{rs}$ are also possible and are observed in real-world situations.

This description does not completely fix the parameters of the model: they are arbitrary to within a multiplicative constant, since one can multiply all the $\theta_i$ in any group by a constant and divide the same constant out of $\omega_{rs}$ without affecting the behavior of the model.  This is why we refer to $\theta_i$ as a ``degree-like parameter''---it plays a role similar to degree in the configuration model, but is arbitrary to within a group-dependent multiplicative constant.  Following~\cite{KN11a}, we remove this ambiguity by making a specific choice of normalization, that the sum of $\theta_i$ within any group should be~1:
\begin{equation}
\sum_j \theta_i \delta_{g_i,r} = 1,
\label{eq:constraint}
\end{equation}
where $\delta_{ij}$ is the Kronecker delta.  This gives us $k$ constraints, one for each of the $k$ groups, and hence fixes all the remaining degrees of freedom.

To generalize this model to the dynamic case we again divide our $n$ nodes into~$k$ groups and assign to each of them a degree-like parameter~$\theta_i$ satisfying~\eqref{eq:constraint}.  We generate an initial state drawn from the static degree-corrected block model with these parameters.  We then generate a history for the network by adding edges between each node pair $i,j$ at rate 
\begin{equation}
\lambda_{ij} = \mu_{rs} \omega_{rs} \theta_i \theta_j
\label{eq:lambdars}
\end{equation}
and removing existing edges independently at rate~$\mu_{rs}$, where $r=g_i$ and $s=g_j$ are respectively the groups to which $i$ and $j$ belong.  Note the similarity between Eqs.~\eqref{eq:lambda} and~\eqref{eq:lambdars}, the primary differences being that the parameter~$\mu_{rs}$ now depends on the group memberships and that the factor~$1/2m$ has been replaced by the quantity~$\omega_{rs}$, which also depends on the group memberships.  By the same argument as before, the number of edges between $i$ and $j$ in the stationary state is Poisson distributed with mean
\begin{equation}
{\lambda_{ij} \over\mu_{rs}} = \omega_{rs} \theta_i \theta_j,
\label{eq:dcsbm}
\end{equation}
which makes the stationary state of this model equivalent to the degree-corrected stochastic block model as desired.

Also by the same argument as before, we can calculate the transition rates for edges to appear and disappear between one snapshot and the next, which are
\begin{align}
\label{eq:transition00}
 p_{0 \to 0} &= \e^{- \beta_{rs} \omega_{rs} \theta_i \theta_j }, \\
 p_{0 \to 1} &= \beta_{rs} \omega_{rs} \theta_i \theta_j  \e^{-\beta_{rs} \omega_{rs} \theta_i \theta_j}, \\ 
 p_{1 \to 0} &= \beta_{rs}\e^{-\beta_{rs} \omega_{rs} \theta_i \theta_j}, \\
\label{eq:transition11}
 p_{1 \to 1} &= (1-\beta_{rs}) \e^{-\beta_{rs} \omega_{rs} \theta_i \theta_j}.
\end{align}
Here
\begin{equation}
\beta_{rs} = 1 - \e^{-\mu_{rs}}
\end{equation}
is the total probability for an existing edge between nodes in groups $r$ and $s$ to disappear in the unit of time between successive snapshots.  (Also as before we have in Eq.~\eqref{eq:transition11} discarded terms beyond leading order in the small quantities~$\omega_{rs}$.)

\begin{widetext}
  By fitting this model to observed network data, we can determine the parameters~$\beta_{rs}$, $\omega_{rs}$, and $\theta_i$, along with the group assignment parameters~$g_i$.  The likelihood as a function function of the four sets of parameters $\set{\beta_{rs}}$, $\set{\omega_{rs}}$, $\set{\theta_i}$, and $\set{g_i}$ takes the form
\begin{equation}
P(\set{\mat{A}(t)}|\set{\beta_{rs}},\set{\omega_{rs}},\set{\theta_i},\set{g_i}) = \prod_{i<j} \biggl[ P(A_{ij}(0)|\beta_{g_ig_j},\omega_{g_ig_j},\theta_i,\theta_j)
  \prod_{t=1}^T P\bigl(A_{ij}(t)| \beta_{g_ig_j},\omega_{g_ig_j},\theta_i,\theta_j,A_{ij}(t-1)\bigr) \biggr].
\label{eq:likelihood}
\end{equation}
The first probability on the right is straightforward, taking the value
\begin{equation}
P(A_{ij}^{(0)}|\beta_{g_ig_j},\omega_{g_ig_j},\theta_i,\theta_j)
  = {(\omega_{g_ig_j} \theta_i \theta_j)^{A_{ij}(0)}\over A_{ij}(0)!}
    \e^{-\omega_{g_ig_j} \theta_i \theta_j}
\end{equation}
by definition (which is independent of~$\beta_{g_ig_j}$), while the second can be expressed in terms of the transition probabilities, Eqs.~\eqref{eq:transition00} to~\eqref{eq:transition11}.  The resulting expression for the log-likelihood is
\begin{align}
\mathcal{L} &= \sum_{ij} \biggl\lbrace A_{ij}(0) \log (\omega_{g_ig_j} \theta_i \theta_j)
  - \omega_{g_ig_j} \theta_i \theta_j
  + \sum_{t=1}^T \biggl[ \bigl[ 1 - A_{ij}(t-1) \bigr] A_{ij}(t)
    \log (\beta_{g_ig_j} \omega_{g_ig_j} \theta_i \theta_j) \nonumber\\
&\hspace{8em}{} + A_{ij}(t-1) \bigl[ 1 - A_{ij}(t) \bigr] \log \beta_{g_ig_j}
  + A_{ij}(t-1) A_{ij}(t) \log(1-\beta_{g_ig_j})
  - \beta_{g_ig_j} \omega_{g_ig_j} \theta_i \theta_j \biggr] \biggr\rbrace \nonumber\\
  &= \sum_{ij} \sum_{rs} \delta_{g_i,r} \delta_{g_j,s}
     \biggl\lbrace A_{ij}(0) \log (\omega_{rs} \theta_i \theta_j)
  - \omega_{rs} \theta_i \theta_j
  + \sum_{t=1}^T \biggl[ \bigl[ 1 - A_{ij}(t-1) \bigr] A_{ij}(t)
    \log (\beta_{rs} \omega_{rs} \theta_i \theta_j) \nonumber\\
&\hspace{8em}{} + A_{ij}(t-1) \bigl[ 1 - A_{ij}(t) \bigr] \log \beta_{rs}
  + A_{ij}(t-1) A_{ij}(t) \log(1-\beta_{rs})
  - \beta_{rs} \omega_{rs} \theta_i \theta_j \biggr] \biggr\rbrace \nonumber\\
  &= \sum_{ij} \biggl[ A_{ij}(0) + \sum_{t=1}^T \bigl[ 1 - A_{ij}(t-1) \bigr]
     A_{ij}(t) \biggr] \log (\theta_i \theta_j)
     + \sum_{rs} \biggl\lbrace m_{rs}(0) \log \omega_{rs} \nonumber\\
&\hspace{8em}{} + m_{rs}^{0\to1} \log (\beta_{rs} \omega_{rs})
  + m^{1\to0}_{rs} \log \beta_{rs} + m^{1\to1}_{rs} \log(1-\beta_{rs})
  - (1 + T \beta_{rs}) \omega_{rs} \biggr] \biggr\rbrace,
\label{eq:loglike}
\end{align}
\end{widetext}
where
\begin{equation}
m_{rs}(0) = \sum_{ij} A_{ij}(0) \delta_{r,g_i} \delta_{s,g_j},
\end{equation}
and
\begin{equation}
m^{0\to1}_{rs} = \sum_{ij} \bigl[ 1 - A_{ij}(t-1) \bigr] A_{ij}(t)
                 \delta_{r,g_i} \delta_{s,g_j},
\end{equation}
which is the total number of edges that appear between groups $r$ and $s$ in the observed data.  Similarly,
\begin{align}
m^{1\to0}_{rs} &= \sum_{ij} A_{ij}(t-1) \bigl[ 1 - A_{ij}(t) \bigr]
                 \delta_{r,g_i} \delta_{s,g_j}, \\
m^{1\to1}_{rs} &= \sum_{ij} A_{ij}(t-1) A_{ij}(t)  \delta_{r,g_i} \delta_{s,g_j},
\end{align}

Differentiating Eq.~\eqref{eq:loglike} with respect to $\omega_{rs}$ now gives~us
\begin{equation}
\omega_{rs} = {m_{rs}(0) + m_{rs}^{0\to1}\over 1+T\beta_{rs}},
\label{eq:omegars}
\end{equation}
and differentiating with respect to~$\beta_{rs}$ gives a quadratic equation again:
\begin{align}
& T\omega_{rs} \beta_{rs}^2 - (T\omega_{rs}
  + m_{rs}^{0\to1} + m_{rs}^{1\to0} + m_{rs}^{1\to1} ) \beta_{rs} \nonumber\\
&\qquad{} + m_{rs}^{0\to1} + m_{rs}^{1\to0} = 0.
\label{eq:betars}
\end{align}
(Note that in order to perform the derivatives correctly, one must take into account the fact that $\omega_{rs}=\omega_{sr}$ and $\beta_{rs}=\beta_{sr}$, although it turns out that the end result is the same as would be derived by naive differentiation, ignoring these equalities.)

Differentiating~\eqref{eq:loglike} with respect to~$\theta_i$, and normalizing appropriately, gives us
\begin{equation}
\theta_i = {\sum_j \bigl\lbrace A_{ij}(0) + \sum_{t=1}^T \bigl[ 1 - A_{ij}(t-1) \bigr] A_{ij}(t) \bigr\rbrace\over\sum_s (1+T\beta_{g_is}) \omega_{g_is}}.
\label{eq:di}
\end{equation}
The self-consistent solution of Eqs.~\eqref{eq:omegars}, \eqref{eq:betars}, and~\eqref{eq:di}, now gives us the parameters of the model.

If we want to convert the degree-like parameter~$\theta_i$ into a true degree, we can do this by noting that the expected degree~$d_i$ of node~$i$ in the stationary state of this model is equal to the sum of the expected number of edges between~$i$ and every other node, which is
\begin{equation}
d_i = \sum_j \omega_{g_ig_j} \theta_i \theta_j
    = \theta_i \sum_{rj} \omega_{g_ir} \theta_j \delta_{g_j,r}
    = \theta_i \sum_r \omega_{g_ir},
\label{eq:di_sbm}
\end{equation}
where we made use of Eq.~\eqref{eq:constraint} in the final equality.  Hence the degrees are simply proportional to~$\theta_i$, with a constant of proportionality that can be easily calculated once we have the values of~$\omega_{rs}$ from Eq.~\eqref{eq:omegars}.

This still leaves us to calculate the maximum-likelihood estimates of the group assignments~$g_i$.  To do this, we substitute our estimates of the parameters back into the log-likelihood, Eq.~\eqref{eq:loglike}, to get the so-called \defn{profile likelihood}, which is then maximized over the group assignments~$g_i$.  Note that there is no need to calculate the last term $\sum_{rs} (1+T\beta_{rs})\omega_{rs}$ in the likelihood since, by Eq.~\eqref{eq:omegars}, it is equal to $\sum_{rs} [ m_{rs}(0) + m_{rs}^{0\to1} ]$, which is independent of the group assignments and hence has no effect on the position of the maximum.

\begin{figure*}[t]
\begin{center}
\includegraphics[width=2\columnwidth]{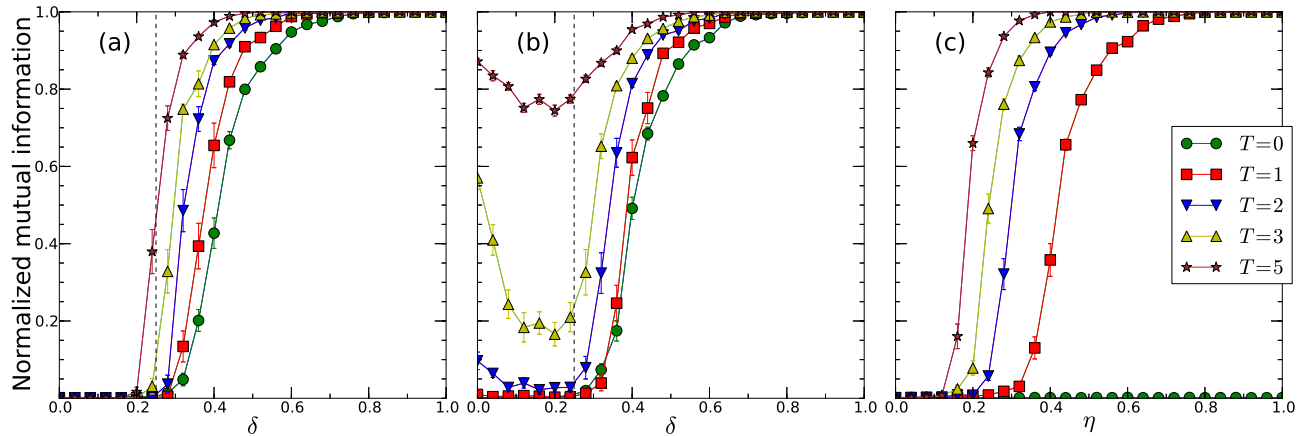}
\end{center}
\caption{The normalized mutual information for runs of the community finding algorithm described here on computer-generated networks themselves created using the dynamic block model.  The parameter~$\delta$ measures the strength of the community structure while $\eta$ measures the extent to which community structure and edge dynamics are correlated.  (a)~Networks with $\eta=0$, $\beta^{\text{uniform}} = 0.4$, and varying $\delta$.  (b)~Networks with $\eta=1$ and $\beta^{\text{planted}}_{rs}$ equal to $\beta_{\text{in}} = 0.3$ along the diagonal and $\beta_{\text{out}}=0.5$ off the diagonal. (c)~Networks with $\delta=0$, $\beta^{\text{uniform}} = 0.4$, $\beta_{\text{in}}=0$, and $\beta_{\text{out}}=0.8$, and varying $\eta$.  The vertical dashed line in panels~(a) and~(b) represents the theoretical detectability threshold for single networks generated from the standard stochastic block model with the same parameters~\cite{DKMZ11a}.  Panel~(b) shows that the dynamics of the network can give us additional information that allows us to find the community structure even below this static threshold.  Each data point is an average over 30 networks with $n=500$ nodes each and average degree $c=16$ for all nodes.}
\label{fig:sbm}
\end{figure*}

Maximization of the profile likelihood over the values of~$g_i$ is harder than maximizing with respect to the other parameters, since the values of the $g_i$ are discrete.  We perform the maximization numerically, using a heuristic algorithm analogous to that used for the static block model in~\cite{KN11a}, which was in turn inspired by the classic Kernighan--Lin algorithm for graph partitioning~\cite{KL70}.  Starting from a random group assignment, we move a single node to a different group, choosing from among all possible such moves the one that most increases (or least decreases) the profile likelihood.  We repeat this process, making a chain of successive single-node moves, but with the important qualification that each node is moved only once.  When all nodes have been moved once, we reexamine every state passed through during the process to find the one with the highest profile likelihood, then take that state as the starting point for a new repetition of the same algorithm.  We continue repeating until no further improvement in the profile likelihood is found.  As with many other optimization algorithms, the results can vary from one run to another because of the random initial condition, so one commonly performs several complete runs with different initial conditions, taking as the final answer the output of the run that gives the highest overall value of the profile likelihood.

An alternative way to fit our model would be to use an expectation--maximization (EM) algorithm in which the model parameters are assigned their maximum-likelihood values but one computes an entire posterior distribution over divisions of the network into groups.  The latter distribution, being a large object, is normally evaluated only approximately, either by Monte Carlo sampling or using a belief propagation algorithm~\cite{DKMZ11a} in which nodes pass each other estimates of their (marginal) probabilities of belonging to each group.  A belief propagation algorithm was used previously for a different dynamic block model in~\cite{Ghasemian15}, where each node sends messages both along ``spatial'' edges to its neighbors in each snapshot and along ``temporal'' edges to its past and future selves in adjacent snapshots.  A similar approach could work in the present case, although our model differs from that of~\cite{Ghasemian15} in assuming unchanging group memberships but correlated edges where~\cite{Ghasemian15} makes the opposite assumption of time-varying group memberships but independent edges between snapshots.

\section{Applications}
\label{sec:examples}
In this section we give examples of fits of dynamic network data to the dynamic configuration model of Section \ref{sec:config} and the dynamic block model of Section~\ref{sec:dsbm}.

\subsection{Synthetic networks}
\label{sec:synthetic}
Our first set of examples make use of synthetic data sets---computer-generated networks with known structure that we attempt to recover using the maximum-likelihood fit.  We demonstrate this approach using the dynamic block model of Section~\ref{sec:dsbm} and the test networks we use are themselves generated using the same model.  We look in particular at the case where the expected degree parameters~$d_i$ for all nodes are the same, equal to a constant~$c$.  For the tests reported here we use $c=16$.  At the same time we varying the strength of the community structure, encapsulated in the parameters~$\omega_{rs}$, according to
\begin{equation}
 \omega_{rs} = \delta \omega_{rs}^{\text{planted}} +(1-\delta) \omega^{\text{random}},
\label{eq:sbm}
\end{equation}
Here $\omega_{rs}^{\text{planted}}$ is diagonal (all elements with $r \ne s$ are zero), $\omega^{\text{random}}$ is a flat matrix (all elements are the same), and $\delta \in [0,1]$ is an interpolating parameter.  Thus by varying $\delta$ we span the range from a uniform random graph with no community structure ($\delta=0$) to a network in which all edges lie within communities and none between communities ($\delta=1$), so that the communities are completely disconnected components.

We similarly vary the rate constants~$\beta_{rs}$ according to a second parameter~$\eta$, also lying in~$[0,1]$, such that
\begin{equation}
 \beta_{rs} = \eta \beta_{rs}^{\text{planted}} +(1-\eta) \beta^{\text{uniform}},
\label{eq:sbm_rho} 
\end{equation}
which interpolates between values that are the same for all groups and the heterogeneous choice~$\beta_{rs}^{\text{planted}}$, which can be anything we choose.  Note that while varying $\beta_{rs}$ does not change the expected degree or average density of edges in the network, it does change how rapidly edges appear and disappear.  Thus $\eta$ controls the extent to which the dynamics of the network, as opposed to merely its average behavior, gives additional information about the community structure.

Once the parameters are fixed, we generate a set of networks, which in our tests have $n=500$ nodes divided into two groups of equal size.  For each network we generate an initial state followed by up to five further snapshots.  The initial state is generated from the stationary distribution (i.e.,~from a traditional degree-corrected block model) and the following snapshots are generated according to the prescription of Section~\ref{sec:dsbm}.

We now apply the fitting method of Section~\ref{sec:dsbm} to these networks to test whether it is able to successfully recover the community structure planted in them.  Success, or lack of it, is quantified using the normalized mutual information~\cite{DDDA05,Meila07}, an information-theoretic metric that measures the agreement between two sets of group assignments.  As traditionally defined, a normalized mutual information of~1 indicates exact recovery of the planted groups while 0 indicates complete failure---zero correlation between recovered and planted values.

Figure~\ref{fig:sbm} shows the results of our tests.  In panel~(a) we fix $\eta=0$, so that $\beta_{rs}$ is uniform and block structure is indicated only by the relative abundance of edges within and between groups.  We use a value of $\beta^{\text{uniform}}=0.4$, meaning that 40\% of extant edges disappear at each time-step.  The different curves in the figure show the normalized mutual information as a function of the parameter~$\delta$ which measures the strength of the community structure, for different numbers of snapshots from $T=0$ to $T=5$.  As we can see, our ability to recover the planted structure diminishes, and eventually fails completely, as the structure becomes weaker, but this effect is partly offset (as we might expect) by increasing the number of snapshots---the more snapshots we use the better we are able to infer the community structure.  For larger numbers of snapshots, the algorithm is able to surpass the known ``detectability threshold'' below which community detection is impossible for single, static networks~\cite{DKMZ11a}, which is indicated by the vertical dashed line in the figure.  In other words the algorithm is able to integrate information about the network over time in order to better determine the shape of the communities.

In Fig.~\ref{fig:sbm}b we set $\eta=1$, so that $\beta_{rs} = \beta^{\text{planted}}_{rs}$, choosing the value of $\beta^{\text{planted}}_{rs}$ to be $\beta_{\text{in}}=0.3$ along the diagonal and $\beta_{\text{out}}=0.5$ off the diagonal, meaning that within-group edges are somewhat more persistent---more likely to be conserved from one snapshot to the next---than between-group edges.  This behavior provides another signal of community structure, in addition to the differing time-averaged edge probabilities, which the algorithm can in principle use to determine group memberships.  And indeed the results of Fig.~\ref{fig:sbm}b reflect this, showing that the algorithm is able to determine group memberships even well below the detectability threshold, but only when $T$ is large.  If $T$ is small, then it becomes difficult to determine the values of $\beta_{rs}$ from the data, and hence difficult to determine group membership for small~$\delta$.  This point is discussed further below.

In Fig.~\ref{fig:sbm}c we fix $\delta=0$ and vary $\eta$ between zero and one using values $\beta^{\text{uniform}}=0.4$ as previously, and $\beta_{\text{in}}=0$, $\beta_{\text{out}}=0.8$.  With $\delta=0$ there is now no signal whatsoever of community structure present in the positions of the edges.  The only clue to the group assignments lies in the rate of appearance and disappearance of edges within and between groups.  As we would expect, the algorithm is unable to identify the communities at all when $T=0$ or $\eta=0$, but as $\eta$ grows for $T\ge1$ the algorithm assigns a larger and larger fraction of nodes to the correct groups, with better performance for larger values of~$T$.  These results suggest the existence of a new detectability threshold as a function of~$\eta$, with location tending to zero as $T\to\infty$.  (A threshold like this was observed, for instance, by Ghasemian~\etal~\cite{Ghasemian15} in their model, discussed in Section~\ref{sec:dsbm}, which has a transition as a function of both the strength of community structure and the relevant rate parameters.)

\begin{figure}
\begin{center}
\includegraphics[width=\columnwidth]{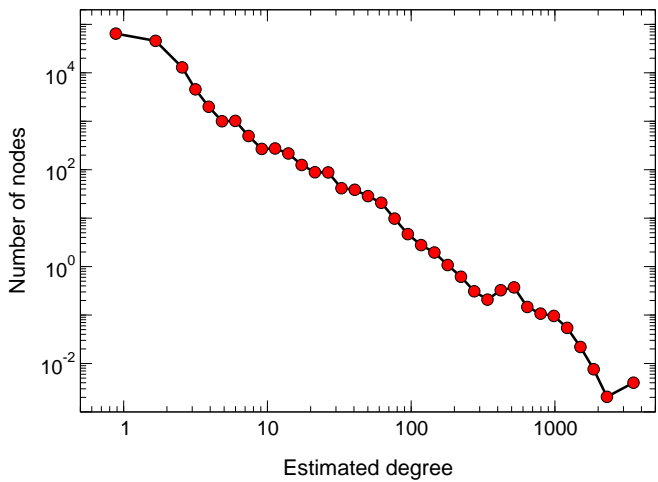}
\end{center}
\caption{Degree distribution of the Internet at the autonomous system level, estimated using the method described here from four snapshots of the network taken at three-month intervals during 2015.  The points are a histogram of estimated degrees using logarithmic (constant ratio) bins.  Note that the expected degrees are not necessarily integers, so the positions of the points are not integers either.}
\label{fig:internet}
\end{figure}

\subsection{Real-world examples}
\label{sec:realworld}
We have also tested our models against a number of empirical data sets representing the structure of real-world dynamic networks.  We give three examples representing networks drawn from technological and social domains, finding in each case that our dynamic models and their associated algorithms perform better than static methods.

\paragraph{Internet graph:}
Our first example is a network representation of the structure of the Internet at the level of autonomous systems (ASes), the fundamental units of global packet routing used by the Internet's Border Gateway Protocol.  The structure of the Internet changes constantly and is well documented: a number of ongoing projects collect snapshots of the structure at regular intervals and make them available for research.  Here we use data from the CAIDA AS Relationships Dataset~\cite{CAIDA}, focusing on four snapshots of the network's structure taken at three-month intervals during 2015.  The spacing of the snapshots is chosen with an eye to the rate of growth of the network.  The Internet has grown steadily in size over the several decades of its existence, and it is still growing today, but this growth is not captured by our models.  To ensure better fits, therefore, we first restrict our data to the set of nodes that are present in all of our snapshots, and second choose snapshots that span a relatively short total time.  Thus our four snapshots were chosen to be sufficiently far apart in time that the network sees significant change between one snapshot and the next, but close enough that the size of the network does not change significantly.

We fit our Internet data to the dynamic version of the configuration model described in Section~\ref{sec:config}, which gives us a way to determine the parameter~$\beta$ that controls the rate of appearance and disappearance of edges as well as the effective degrees~$d_i$ of nodes~$i$ in the network.  For the rate parameter we find a maximum likelihood value of $\beta=0.0896$, which indicates a fairly slow rate of turnover of the edges in the network.  Recall that $\beta$ is the average probability that an edge will vanish from one snapshot to the next, so this value of $\beta$ implies that over 90\% of edges remain intact between snapshots.  As discussed in Section~\ref{sec:config}, one could make a naive estimate of the rate at which edges vanish simply by counting the number that do, but that estimate would be less accurate than the maximum-likelihood one.

Our fit also gives us estimates of the degree parameters~$d_i$ from Eq.~\eqref{eq:di_config}.  Figure~\ref{fig:internet} shows a histogram of the frequency distribution of estimated degrees for the Internet derived in this manner.  Again, we could make naive estimates of the degrees, for instance by assuming snapshots to be independent and averaging the raw degrees of their nodes across snapshots.  This would be a correct estimator of the~$d_i$ in the limit of a large number of snapshots, meaning it will tend to the correct answer eventually, but it would be less than ideal.  In particular, our estimate of the error on the values it gives would be wrong.  By assuming the snapshots to be independent, we effectively assume that we have more measurements than we really do and hence underestimate the variance.  For instance, if we observe that the naive degree of a node is unchanging for many snapshots in a row, we may conclude that the average of those values has a very small statistical error, because the fluctuations are small.  This, however, would be erroneous if the small fluctuations are actually just a result of the fact that the network is only changing rather slowly.

Error estimates are not the only thing that will be affected by improperly using a naive degree estimate.  The values of the degrees themselves can also be affected if the snapshots are strongly correlated, which they are in this case because of the small value of~$\beta$.  Strongly correlated snapshots will tend to give a node the same or similar degree on successive snapshots, but Eq.~\eqref{eq:di_config} implies that in this case our estimate of~$d_i$ should actually \emph{decrease} over time (as $T$ becomes larger in the denominator while the numerator remains constant).  A naive estimate on the other hand would remain unchanged.  At first sight the decrease in the maximum-likelihood estimate may appear counterintuitive, but it has a simple physical interpretation: for a node that truly has a constant value of~$d_i$, we would expect additional edges to appear occasionally, at a rate dependent on the value.  If we do not see any edges appearing, therefore, it implies our initial estimate of the degree was too high and we should revise it downward.

The maximum-likelihood estimator can, on the other hand, also have problems of its own if the the model we are fitting is not a perfect description of the data.  In the case of the Internet we see two possible sources of disagreement between data and model.  First, even though the number of nodes in the network is held fixed, the number of edges is observed to grow over time---the network is becoming more dense.  This effect is not included in our model, which assumes constant expected density.  Second, we see some evidence that the removal of edges is not uniform as our model assumes, but that edges connected to high-degree nodes disappear at a higher rate than those connected to low-degree ones.  Both of these behaviors could potentially affect our results.  (It is interesting to ask whether and how the model could be generalized to include them, though we leave pursuit of these questions for future work.)

\paragraph{Friendship network:} Our second example focuses on a set of social networks from a study by Michell and West of friendship patterns and behaviors among school students in the UK~\cite{MW96}.  High-school students at a school in the west of Scotland were polled about their friendship patterns, each student being allowed to name up to twelve friends, and they were also asked about their drinking, smoking, and drug use habits.  The entire exercise was conducted a total of three times, at yearly intervals, with the same group of students.  The study looked at all students in the school, but the most detailed data were collected for a subset of 50 girls within the larger population and it is on this subset that we focus here.

The researchers were interested in the extent to which substance use behaviors correlated with friendship patterns.  They found that although there was no single factor that would completely explain the friendships of the students, the network of friendships did display homophily according to substance use, meaning that students with similar use patterns were more likely to be friends~\cite{PW03,PSS06}.

\begin{figure}
\begin{center}
\includegraphics[width=\columnwidth,clip=true]{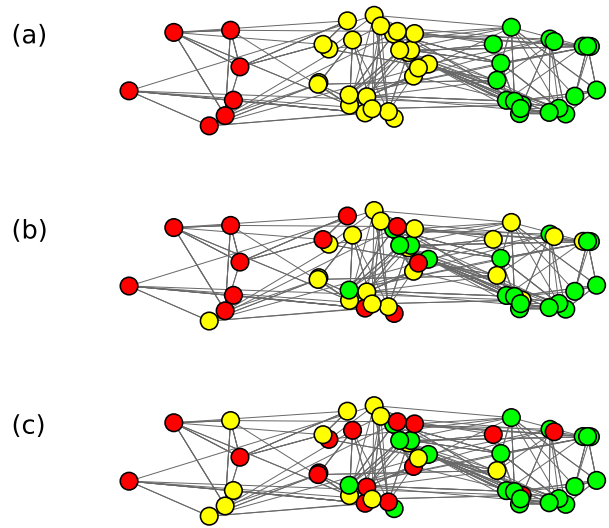}
\end{center}
\caption{Communities within the friendship network of UK high-school students described in the text.  (a)~Groups are colored according to ground-truth data on substance use, where the colors from green to red indicate students who used zero, one, or two or more substances, respectively.  (b)~Colors indicate group assignments inferred by fitting the network to the dynamic block model of this paper using all three snapshots.  (c)~Colors indicate the group assignments inferred by fitting an aggregate of the three snapshots to the static degree-corrected stochastic block model.}
\label{fig:scotland}
\end{figure}

In our analysis, we divide the students into three groups: those who do not drink, smoke, or take drugs on a regular basis; those who exhibit one of these three behaviors; and those who exhibit two or more.  We then ask whether it is possible to detect this division into groups based on network structure alone, without any knowledge of student behaviors.  We find that when using the dynamic version of the degree-corrected block model described in Section~\ref{sec:dsbm} it is indeed possible to determine the groups, and to do so with better accuracy than can be achieved by standard static methods.  Specifically, we compare results from our dynamic block model to those from the static degree-corrected block model fitted to an aggregate network formed from the union of the three snapshots.

Figure~\ref{fig:scotland} shows three pictures of the overall aggregate network of friendships.  Each picture is laid out identically, but with different coloring.  In panel~(a) the three colors represent the ground truth, with green, yellow, and red denoting students who engaged in zero, one, or two or more of the behaviors studied respectively.  Panel~(b) shows the communities found in the network by fitting to the dynamic block model.  Though not perfect, this fit places 64\% of the nodes in their correct groups.  A random coloring, for comparison, would get only 33\% right.  Panel~(c) shows the results from the standard static algorithm applied to the aggregated network.  This fit places only 52\% of the nodes in their correct groups.

\begin{figure}
\begin{center}
\includegraphics[width=6cm,clip=true]{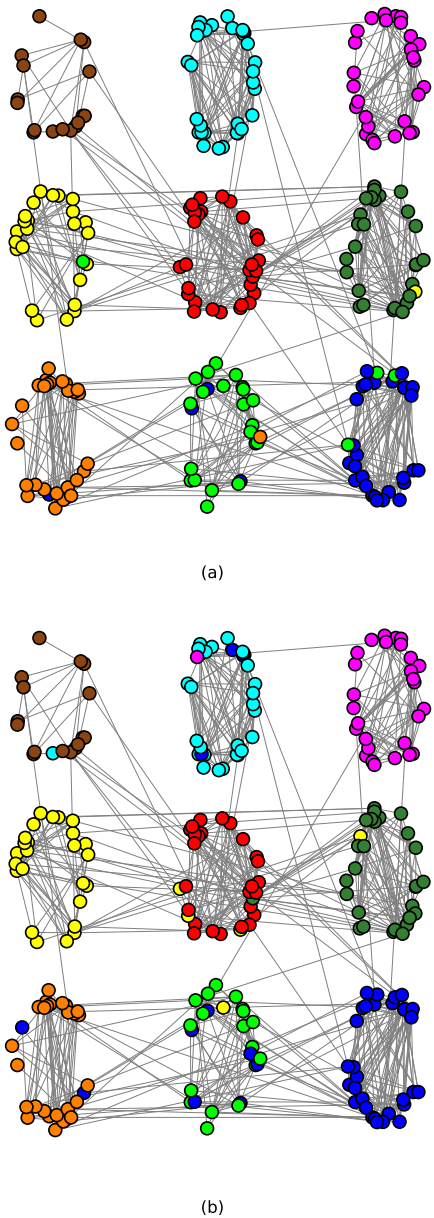}
\end{center}
\caption{Student proximity network.  The nine groups of nodes in each panel represent the nine classes and the colors represent the community structure found using (a)~the dynamic model of this paper and (b)~the standard static degree-corrected block model applied to the aggregate of all four snapshots.  Classes in the same row belong to the same subject specialty and tend to have more inter-class edges than classes in different rows.}
\label{fig:france}
\end{figure}

\paragraph{Proximity network:} Our third example is another social network, a network of physical proximity between students in a high school in France~\cite{MFB15}.  The data were collected using electronic proximity detectors worn by the participants.  The detectors record the presence and identity of other detectors in their vicinity at intervals of 20 seconds.  The data were collected over five consecutive days, but only a half day's worth of data were collected on the last day, which we discard, leaving four full days to work with.  We construct one snapshot for each day and consider there to be an edge between two participants in a snapshot if three or more contacts between them are recorded during the relevant day.  Requiring a minimum number of contacts in this way helps to remove spurious signals from the data, as discussed in~\cite{SBPT16}.  We also restrict our study to those nodes that are present in all snapshots.

The students in the study were divided among three subject specialties: mathematics/physics, physics/chemistry, and engineering.  Each specialty was further divided into three classes, so there are a total of nine classes in the data.  We attempt to recover these classes from the network data alone, without other information, using both the dynamic model of this paper and a traditional static degree-corrected block model applied to the aggregated network.  In this case both methods do well, which is perhaps unsurprising, given that the edges within each group are significantly denser than those between groups.  Figure~\ref{fig:france} shows the results for the dynamic model in panel~(a) and the static model in panel~(b).  As we can see, both models achieve good classification of the nodes into their classes, though the dynamic model performs slightly better.  The error rate---the fraction of incorrectly labeled nodes---is 4.1\% for the dynamic model of panel~(a) and 5.7\% for the static model of panel~(b).

The primary benefit of the dynamic model in this case, however, lies not in its ability to recover the communities but in what it reveals about the dynamics of the network.  In addition to the communities themselves, the dynamic model also returns values for the rate parameters that can reveal features of the data not seen in the simple static fit to the aggregate network.  Of particular interest in this case are the parameters~$\beta_{rs}$, which measure the relative rates at which edges change within and between groups.  Our fit gives estimates of
\begin{equation}
 \beta_{rs} \simeq
\begin{cases}
 0.51 \quad \text{within classes,} \\
 0.75 \quad \text{different classes but the same specialty,} \\
 0.94 \quad \text{different specialties.} 
\end{cases}
\end{equation}
In other words, connections are not only more likely between participants in the same class or specialty, but they are also more persistent, in some cases by a wide margin---only about 6\% of connections persist from one snapshot to the next between individuals in the different specialties for example, but almost 50\% persist within classes.  (There is some variation in values of~$\beta_{rs}$ among classes and specialties; the results above are only an approximate guide based on average values for each type.)

\section{Conclusions}
\label{sec:conclusions}
In this paper we have introduced dynamic generalizations of some of the best-known static network models, including the Erd\H{o}s--R\'enyi random graph, the configuration model, and the degree-corrected stochastic block model.  We have also derived and implemented efficient algorithms for fitting these models to network data that allow us to infer maximum-likelihood estimates of rates of change, node degrees, and community structure.  We have tested the performance of our models and algorithms on synthetic benchmark networks as well as on a selection of data sets representing real-world examples of dynamic networks.

There are a number of directions in which this work could be extended.  First, we have focused exclusively on edge dynamics here, but there are also networks in which nodes appear and disappear and it would be a natural generalization to study the dynamics of nodes also, or of both edges and nodes together.  (We could also  allow node properties, such as expected degrees or community memberships, to change over time, as some other authors have done.)  Second, the assumption of continuous-time Markov processes for the edge dynamics is a particularly simple one, which could be relaxed to encompass more complicated situations.  Third, in our community detection calculations we assume we know the number of communities the network contains, but in many cases we do not have this information.  Methods have been developed for determining the number of communities in static networks and it is an interesting question whether those methods can be extended to the dynamic case as well.

\begin{acknowledgments}
  The authors thank Aaron Clauset for useful conversations.  This research was supported in part by the US National Science Foundation under grants DMS-1107796 and DMS-1407207 (MEJN) and by the Army Research Office under grant W911NF-12-R-0012 and the John Templeton Foundation (CM).
\end{acknowledgments}

\end{document}